\documentclass[a4paper,twoside,11pt]{article} 

\usepackage[english]{babel} %
\usepackage{lmodern}
\usepackage[T1]{fontenc}
\usepackage[latin9]{inputenc}	

\newlength{\defaultparindent}
\usepackage{latexsym}
\usepackage{amsmath}
\usepackage{amsfonts}
\usepackage{amsthm}
\usepackage{bbold}
\usepackage{multirow}
\usepackage{hyperref}
\hypersetup{colorlinks=true,citecolor=green,linkcolor= blue}
\usepackage{todonotes} 
\usepackage{soul} 
\usepackage{tikz-cd}

\usepackage{algorithm} 
\usepackage{algpseudocode}

\usepackage[arXiv]{optional} 
\def\mynote{\todo} 

\opt{AACA}{
\def\cal{\mathcal}
}

\opt{x,AACA}{
\input{mbh_defs_TeX}
}

\opt{arXiv,std,JMP,JOPA}{

\newtheorem{MS_theorem}{Theorem}
\newtheorem{MS_lemma}{Lemma}
\newtheorem{MS_Proposition}{Proposition}
\newtheorem{MS_Corollary}[MS_Proposition]{Corollary}

\def\eg{e.g.\ }
\def\myisom{\cong} 
\newcommand{\R}{\ensuremath{\mathbb{R}}} 
\newcommand{\Identity}{\ensuremath{\mathbb{1}}} 
\newcommand*\sepline{%
  \begin{center}
    \rule[1ex]{.3\textwidth}{.1pt}
  \end{center}}

\def\my_span#1{\mbox{Span}\left(#1\right)} 

\def\bino#1#2{#1 \choose #2} 
\def\dotinformula{\;\; \mathrm{.}} 
\def\OO#1{\ensuremath{\mbox{O}\!\left(#1\right)}}
\def\SO#1{\ensuremath{\mbox{SO}\!\left(#1\right)}}
\def\GL#1{\ensuremath{\mbox{GL}\!\left(#1\right)}}

\def\End{\ensuremath{\mbox{End}}}
\def\O1#1{\ensuremath{\mbox{O}^{#1}(1)}}

\newcommand{\anticomm}[2]{\ensuremath{\left\{ #1, #2 \right\}}} 
\newcommand{\myClg}[3]{\ensuremath{{{\cal C}\ell} {\left( #3 \right)}}}	

}

\opt{JMP}{
}

\def\h_eigen{\eta}
\def\g_eigen{\theta}
\def\mygen{e} 
\def\myseparation{\sepline} 

\def\SAT{\ensuremath{\mbox{SAT}}}

\def\mysetM{\ensuremath{{{\cal M}_n}}} 
\def\mysnqG{\ensuremath{{{\cal N}_n}}} 

\begin{document}

\opt{x,std,arXiv,JMP,JOPA}{
\title{{\bf On Simple Spinors, Null Vectors and the Orthogonal Group $\OO{n}$} 
	}

\author{\\
	\bf{Marco Budinich}%
%
%
\\
	University of Trieste and INFN, Trieste, Italy\\ 
	\texttt{mbh@ts.infn.it}\\
%
%
%
\\	March 31, 2020, submitted
	}
\date{ } 
\maketitle
}

\vspace*{-5mm}
\begin{abstract}
We explore the three separate isomorphisms that link together simple spinors, null vectors and the orthogonal group $\OO{n}$ and exploit them to look back at these arguments from a unified viewpoint.
\end{abstract}


\opt{x,std,arXiv,JMP,JOPA}{
{\bf Keywords:} {Clifford algebra; orthogonal group; involutions.}
}

\opt{AACA}{
\keywords{Clifford algebra; orthogonal group; involutions.}
\maketitle
}

\section{Introduction}
\label{sec_Introduction}
\opt{margin_notes}{\mynote{mbh.note: for paper material see log pp. 714 ff}}%
We explore here the connections that link three different topics: the simple spinors of the Clifford algebra of neutral space $\R^{n,n}$, the null vectors of $\R^{n,n}$ and the orthogonal group $\OO{n}$. The three subjects are tightly intertwined and some of the pivotal results in this field date back
\opt{margin_notes}{\mynote{mbh.note: solita frasetta su Cartan ?? to more than a cantury ago... Un sentiero poco battuto è quello...}}%
to the work of Élie Cartan \cite{Cartan_1937} but, after that, rarely surfaced again in the literature.

These connections are central to highlight the role of Clifford algebra as a possible bridge between the continuous and discrete worlds \cite{Budinich_2018_a} and our goal here is to try to consolidate this bridge and in particular to sustain the application of Clifford algebra to Boolean SATisfiability (\SAT{}) problems \cite{Budinich_2019}.

In section~\ref{sec_neutral_R_nn} we succintly introduce the Clifford algebra $\myClg{}{}{\R^{n,n}}$ of neutral space $\R^{n,n}$ and the formalism we need, in section~\ref{sec_O(n)_MTNP} we review the isomorphism between the orthogonal group $\OO{n}$ and the set \mysnqG{} of all null subspaces of maximal dimension $n$ of $\R^{n,n}$ or Maximally Totally Null Planes, MTNP for short. In following sections~\ref{sec_clauses_isometries} and \ref{sec_Witt_decomposition} we focus on this isomorphism when restricted to an Abelian subgroup of $\OO{n}$ and show how it can be extended also to the set of Witt decompositions of $\R^{n,n}$ and define ``orthogonal'' Witt decompositions. Section~\ref{sec_special_property} contains a constructive proof of a particular property of set \mysnqG{} of $\R^{n,n}$, first proved by Cartan \cite{Cartan_1937} and rarely appearing in subsequent papers: in a nutshell this result proves that the set \mysnqG{}, isomorphic to the orthogonal group $\OO{n}$, is {\em discrete} and contains $2^n$ different MTNP. This is somewhat surprising since MTNP are linear subspaces of $\R^{n,n}$ and is the cornerstone on which the hypothetical bridge between the continuous and discrete worlds would stand. In section~\ref{sec_different_forms_MTNP} we present explicitly the different forms $t \in \OO{n}$ that the same MTNP may assume and show that all together these forms give a cover of the group $\OO{n}$ while in section~\ref{sec_Simple_spinors_MTNP} we close the circle and, exploiting the isomorphism between simple spinors and MTNP, we bring these results in the spinor space of Clifford algebra giving another geometrical interpretation to simple spinors. Finally, in section~\ref{sec_Conclusions}, we try to wrap up all these relations with the aid of a couple of commutative diagrams.

For the convenience of the reader we tried to make this paper as elementary and self-contained as possible for example using, whenever possible, the universal matrix formalism of the vectorial representation of $\OO{n}$.

\section{$\R^{n,n}$ and its Clifford algebra}
\label{sec_neutral_R_nn}
We review here some properties of the neutral space $\R^{n,n}$ and of its Clifford algebra $\myClg{}{}{\R^{n,n}}$ that are at the heart of the following results.

$\myClg{}{}{\R^{n,n}}$ is isomorphic to the algebra of real matrices $\R(2^n)$ \cite{Porteous_1981} and this algebra is more easily manipulated exploiting the properties of its Extended Fock Basis (EFB, see \cite{Budinich_2016} and references therein) with which any algebra element is a linear superposition of simple spinors. The $2 n$ generators of the algebra $\mygen_{i}$ form an orthonormal basis of the neutral vector space $\R^{n,n}$
\begin{equation}
\label{formula_generators}
\mygen_i \mygen_j + \mygen_j \mygen_i := \anticomm{\mygen_i}{\mygen_j} = 2 \left\{ \begin{array}{l l}
\delta_{i j} & \mbox{for} \; i \le n \\
- \delta_{i j} & \mbox{for} \; i > n
\end{array} \right.
\qquad i,j = 1,2, \ldots, 2 n
\end{equation}
and we define the Witt, or null, basis of $\R^{n,n}$:
\begin{equation}
\label{formula_Witt_basis}
\left\{ \begin{array}{l l l}
p_{i} & = & \frac{1}{2} \left( \mygen_{i} + \mygen_{i + n} \right) \\
q_{i} & = & \frac{1}{2} \left( \mygen_{i} - \mygen_{i + n} \right)
\end{array} \right.
\quad i = 1,2, \ldots, n
\end{equation}
that, with $\mygen_{i} \mygen_{j} = - \mygen_{j} \mygen_{i}$ for $i \ne j$, gives
\begin{equation}
\label{formula_Witt_basis_properties}
\anticomm{p_{i}}{p_{j}} = \anticomm{q_{i}}{q_{j}} = 0
\qquad
\anticomm{p_{i}}{q_{j}} = \delta_{i j}
\end{equation}
showing that all $p_i, q_i$ are mutually orthogonal, also to themselves, that implies $p_i^2 = q_i^2 = 0$ and are thus null vectors.
\opt{margin_notes}{\mynote{mbh.note: should I mention here $\R^{2 n}_{hb}$ ? See book 780}}%
Defining
\begin{equation}
\label{formula_Witt_decomposition}
\left\{ \begin{array}{l}
P = \my_span{p_1, p_2, \ldots, p_n} \\
Q = \my_span{q_1, q_2, \ldots, q_n}
\end{array} \right.
\end{equation}
it is easy to verify that they form two MTNP, namely totally null subspaces of maximun dimension $n$; $P$ and $Q$ constitute a Witt decomposition \cite{Porteous_1981} of the neutral space $\R^{n,n}$ since $P \cap Q = \{ 0 \}$ and $P \oplus Q = \R^{n,n}$.
\opt{margin_notes}{\mynote{mbh.note: commented there is a shorter form}}%
%

The $2^{2 n}$ simple spinors forming EFB are given by all possible sequences
\begin{equation}
\label{EFB_def}
\psi = \psi_1 \psi_2 \cdots \psi_n \qquad \psi_i \in \{ q_i p_i, p_i q_i, p_i, q_i \} \qquad i = 1, 2, \ldots, n \dotinformula
\end{equation}

\noindent Any element of (\ref{EFB_def}) represents a simple spinor, and thus an element of one of the $2^n$ Minimal Left Ideals (MLI) of $\myClg{}{}{\R^{n,n}}$; each of these MLI is a spinor space $S$, uniquely identified in EFB by its $h \circ g$ signature \cite{Budinich_2016}, and they are equivalent%
\opt{margin_notes}{\mynote{mbh.ref: Porteous 1995 p. 133 B\&T p. ? 
}}%
{} in the sense that each of them can carry a regular representation of the algebra; moreover the algebra, as a vector space, is the direct sum of these spinor spaces. In the isomorphic matrix algebra $\R(2^n)$ each spinor space $S$ is usually one of the $2^n$ columns of the matrix.

For each of these $2^n$ MLI its $2^n$ simple spinors (\ref{EFB_def}), uniquely identified in EFB by their $h$ signature \cite{Budinich_2016}, form a Fock basis $F$ of the spinor space $S$ and any spinor $\Psi \in S$ is a linear combination of the simple spinors $\psi \in F$ \cite{BudinichP_1989, Budinich_2016}. We illustrate this with the simplest example in $\R^{1,1}$, the familiar Minkowski plane of physics, here $\myClg{}{}{\R^{1,1}} \myisom \R(2)$ and the EFB (\ref{EFB_def}) is formed by just 4 elements: $\{ qp_{+ +}, pq_{- -}, p_{- +}, q_{+ -} \}$ with the subscripts indicating respectively $h$ and $h \circ g$ signatures that give the binary form of the integer matrix indexes; its EFB matrix is
\begin{equation*}
\bordermatrix{& + & - \cr
+ & q p & q \cr
- & p & p q \cr }
\end{equation*}
and, as anticipated, we can write the generic element $\mu \in \myClg{}{}{\R^{1,1}}$ in EFB
$$
\mu = \xi_{+ +} qp_{+ +} + \xi_{- -} pq_{- -} + \xi_{- +} p_{- +} + \xi_{+ -} q_{+ -} \qquad \xi \in \R \dotinformula
$$
The two MLI are the two columns that are two (equivalent) spinor spaces $S_+$ and $S_-$. The two elements of each column are the simple spinors of the Fock basis $F$.

In turn the simple spinors of Fock basis $F$ are in one to one correspondence with MTNP \cite{BudinichP_1989}, for any $\psi \in F$ we define its associated MTNP $M(\psi)$ as
\begin{equation}
\label{formula_MTNP_M_psi}
M(\psi) = \my_span{x_1, x_2, \ldots, x_n} \quad x_i = \left\{ \begin{array}{l l}
p_i & \mbox{iff} \; \psi_i = p_i, p_i q_i \\
q_i & \mbox{iff} \; \psi_i = q_i, q_i p_i
\end{array} \right.
\quad i = 1,2, \ldots, n
\end{equation}
\noindent and $x_i$ is determined by the $h$ signature of $\psi$ in EFB \cite{Budinich_2016}, for example in $\myClg{}{}{\R^{3,3}}$ the simple spinor $\psi = p_1 q_1 \, q_2 p_2 \, q_3 p_3$, of $h$ signature $-++$, is annihilated by any vector of MTNP $M(\psi)$
$$
\psi = p_1 q_1 \, q_2 p_2 \, q_3 p_3 \quad \implies \quad M(\psi) = \my_span{p_1, q_2, q_3}
$$
and it is simple to verify with (\ref{formula_Witt_basis_properties}) that for any $v \in M(\psi)$ then $v \psi = 0$. We will prove more precisely this correspondence in section~\ref{sec_Simple_spinors_MTNP}.

\section{The orthogonal group $\OO{n}$ and the set of MTNP}
\label{sec_O(n)_MTNP}
Let \mysetM{} be the set of $2^n$ MTNP $M(\psi)$ (\ref{formula_MTNP_M_psi}) associated to the $2^n$ simple spinors $\psi \in F$; they are the $2^n$ MTNP formed as the span of the $n$ null vectors obtained choosing one element from each of the $n$ couples $(p_i, q_i)$ (\ref{formula_Witt_basis}) \cite{BudinichP_1989, Budinich_2016}.

\mysetM{} is a subset of the larger set \mysnqG{} of all MTNP of $\R^{n,n}$, for Ian Porteous a semi-neutral quadric Grassmannian \cite{Porteous_1981}. \mysnqG{} in turn is isomorphic to the subgroup $\OO{n}$ of $\OO{n,n}$, moreover $\OO{n}$ acts transitively on \mysnqG{} and thus also on \mysetM{}.

We review these relations: seeing the neutral space $\R^{n,n}$ as $\R^n \times \R^n$ we can write its generic element as $(x,y)$ and then $(x,y)^2 = x^2 - y^2$ and so for any $x \in \R^n \times \{0\}$ and $t \in \OO{n}$ $(x,t(x))$ is a null vector since $(x,t(x))^2 = x^2 - t(x)^2 = 0$ and thus as $x$ spans the (spacelike) subspace $\R^n \times \{0\}$ then $(x,t(x))$ spans a MTNP of $\R^{n,n}$; we indicate this MTNP with self-explanatory notation as $(\Identity, t)$. Isometry $t \in \OO{n}$ establishes the quoted isomorphism since any MTNP of \mysnqG{} can be written in the form $(\Identity, t)$ \cite[Corollary~12.15]{Porteous_1981} and thus the one to one correspondence between MTNP of $\R^{n,n}$ and $t \in \OO{n}$ is formally established by the map
\opt{margin_notes}{\mynote{mbh.note: is this the Cayley chart ?}}%
\begin{equation}
\label{formula_bijection_Nn_O(n)}
\OO{n} \to \mysnqG; t \to (\Identity, t) \qquad \implies \qquad \mysnqG = \{ (\Identity, t) : t \in \OO{n} \} \dotinformula
\end{equation}

For example in this setting the form of two generic null vectors of $P$ and $Q$ (\ref{formula_Witt_decomposition}) are respectively $(x, x)$ and $(x, -x)$ and in our notation we represent the whole MTNP $P$ and $Q$ (\ref{formula_Witt_decomposition}) with
\opt{margin_notes}{\mynote{mbh.ref: see pp. 55, 56}}%
\begin{equation}
\label{formula_P_Q_def}
\left\{ \begin{array}{l}
P = (\Identity, \Identity) \\
Q = (\Identity, -\Identity) \dotinformula
\end{array} \right.
\end{equation}
The action of $\OO{n}$ is transitive on \mysnqG{} since for any $t, u \in \OO{n}$, $(\Identity,ut) \in \mysnqG$ and the action of $\OO{n}$ is trivially transitive on $\OO{n}$.

\section{Discrete isometries of $\OO{n}$}
\label{sec_clauses_isometries}
We examine bijection (\ref{formula_bijection_Nn_O(n)}) when restricted to the subset $\mysetM \subset \mysnqG$ and we take $P = (\Identity, \Identity)$ (\ref{formula_P_Q_def}) as our ``reference'' MTNP of \mysetM{}.
\opt{margin_notes}{\mynote{mbh.note: in physics language this corresponds to the choice of the vacuum spinor}}%

Let $\lambda_i$ be the hyperplane reflection inverting timelike vector $\mygen_{i + n}$, namely
\begin{equation}
\label{t_i_def}
\lambda_i(\mygen_{j}) =
\left\{ \begin{array}{l l l}
-\mygen_{j} & \quad \mbox{for} \quad j = i + n\\
\mygen_{j} & \quad \mbox{otherwise}
\end{array} \right.
\quad i = 1,2, \ldots, n \quad j = 1,2, \ldots, 2n
\end{equation}
its action on the Witt basis (\ref{formula_Witt_basis}) is to exchange the null vectors $p_i$ and $q_i$. It follows that the inversion of a certain subset of timelike vectors exchanges the corresponding null vectors $p_i$ with $q_i$ and viceversa. It is thus clear that starting from $P$ we can obtain any element of \mysetM{} by the corresponding inversion of a subset of the $n$ timelike vectors $\mygen_{i + n}$. Each of these isometries acts on the (timelike) subspace $\{0\} \times \R^n$ of $\R^{n,n}$ and is represented, in the vectorial representation of $\OO{n}$, by a diagonal matrix $\lambda \in \R(n)$ with $\pm 1$ on the diagonal and these matrices form the group $\OO{1} \times \OO{1} \cdots \times \OO{1} = \stackrel{n}{\times} \OO{1} := \O1{n}$ that is immediate to get since $\OO{1} = \{ \pm 1 \}$.%
\opt{margin_notes}{\mynote{mbh.ref: remember that the tensor product is associative Proposition~11.5 of {Porteous 1995}}}%
{} \O1{n} is a discrete, abelian, subgroup of $\OO{n}$. It is thus clear that since $P = (\Identity, \Identity)$ for any $R \in \mysetM$ there exists a unique $\lambda \in \O1{n}$ giving
$$
R = (\Identity, \lambda)
$$
and thus we proved constructively
\begin{MS_Proposition}
\label{isomorphism_restricted}
The isomorphism (\ref{formula_bijection_Nn_O(n)}) when restricted to the subgroup \O1{n} of $\OO{n}$ has for image $\mysetM \subset \mysnqG$
\begin{equation}
\label{formula_bijection_Mn_O^n(1)}
\O1{n} \to \mysetM; \lambda \to (\Identity, \lambda) \qquad \implies \qquad \mysetM = \{ (\Identity, \lambda) : \lambda \in \O1{n} \} \dotinformula
\end{equation}
\end{MS_Proposition}

Subgroup \O1{n} is connected to involutions, namely linear maps $t \in \End \R^n$ such that $t^2 = \Identity$, given that any $\lambda \in \O1{n}$ is an involution of $\R^n$, in particular \O1{n} is a group of commuting involutions. We remind that any involution $t$ splits the linear space into two orthogonal subspaces: the elements such that $t(x) = x$ and those such that $t(y) = -y$, moreover:
\begin{MS_Proposition}
\label{prop_involutions}
Any linear involution $t$ is in $\OO{n}$ and is of the form $t = E \lambda E^T$ for $E \in \OO{n}$ and $\lambda \in \O1{n}$.
\end{MS_Proposition}
\begin{proof}
Let $t = E \lambda E^T$ with $E \in \OO{n}$ and $\lambda \in \O1{n}$, then $t^2 = \Identity$ since $\lambda^2 = \Identity$; conversely if $t$ is an involution $t^2 = \Identity$ implies $t \in \OO{n}$
\opt{margin_notes}{\mynote{mbh.ref: for $t^2 = \Identity \implies t \in \OO{n}$ see log. p. 699'}}%
and then $t = t^T$ and so $t$ has real eigenvalues and $t = E \lambda E^T$ with $\lambda \in \O1{n}$.
\end{proof}


\section{On Witt decompositions of $\R^{n,n}$}
\label{sec_Witt_decomposition}
\opt{margin_notes}{\mynote{mbh.note: see log pp. 716 ff}}%
We begin with two lemmas proving basic MTNP properties:
\begin{MS_lemma}
\label{MTNP_equivalence}
Given any $u, s \in \GL{n}$ such that $s u^{-1} \in \OO{n}$ then $(u, s)$ and $(\Identity, s u^{-1})$ represent the same MTNP of $\R^{n,n}$.
%
\opt{margin_notes}{\mynote{mbh.note: very similar to \cite[Proposition~12.14]{Porteous_1981}, for a fully general version see log. p. 721'}}%
\end{MS_lemma}
\begin{proof}
Since $u, s \in \GL{n}$ any $(u(a), s(a)) := (x, y)$ may be written as $(x, s u^{-1}(x))$, namely $(\Identity, s u^{-1})$, that, by (\ref{formula_bijection_Nn_O(n)}), is a MTNP.
\end{proof}
%
%
%
\noindent 
It follows that an abstract MTNP may assume different equivalent forms: with $(\Identity, t) \simeq (\Identity, u)$ we will mean that the two forms represent the same MTNP; we remark also that the definition makes sense only if $t$ and $u$ are expressed in the \emph{same} basis since it is obvious that any $t \in \OO{n}$ may take the form of any other $\OO{n}$ element for a suitable, orthogonal, change of basis. Moreover this shows also that not only all the vectors defining MTNP $(\Identity, t)$ with $t \in \OO{n}$ are reciprocally orthogonal like in (\ref{formula_Witt_basis_properties}) but also that all of their linear combinations, \eg $(s, t s)$ for any $s \in \GL{n}$, remain reciprocally orthogonal, a property that holds only for MTNP subspaces and that is at the heart of following results.

Two MTNP $R, R' \in \mysnqG$ form a Witt decomposition of $\R^{n,n}$ iff $R \oplus R' = \R^{n,n}$ and since by (\ref{formula_bijection_Nn_O(n)}) we can represent any MTNP as $(\Identity, t)$ we will indicate any couple of MTNP as column vectors by $W = \left(\begin{array}{r r} \Identity & \Identity \\ t & t' \end{array}\right)$ looking at the column vectors of $t$ as at the action of $t$ on the standard basis $\Identity$ given that $t \Identity = t$.

\begin{MS_lemma}
\label{MTNP_diagonalization}
Given any two MTNP forming $W = \left(\begin{array}{c c} \Identity & \Identity \\ t_1 & t_2 \end{array}\right)$ it is always possible to rearrange the basis in $\R^{n, n}$ so that the null vectors forming $W$ take the form $W' = \left(\begin{array}{c c} \Identity & \Identity \\ \Identity & u^{-1} t u \end{array}\right)$ with $t = t_1^T t_2$ and for any $u \in \GL{n}$.
\end{MS_lemma}
\begin{proof}
The vectors of $W$ in basis $B = \left(\begin{array}{c c} u & 0 \\ 0 & t_1 u \end{array}\right)$
take the form $W' = B^{-1} W = \left(\begin{array}{c c} u^{-1} & u^{-1} \\ u^{-1} & u^{-1} t_1^T t_2 \end{array}\right)$ and the thesis is proved applying Lemma~\ref{MTNP_equivalence}.
\end{proof}

\noindent This shows that: i) any MTNP $(\Identity, t_1)$ may be chosen to be the reference MTNP $P = (\Identity, \Identity)$%
\footnote{corresponding to the freedom of choosing an arbitrary vacuum spinor in $S$}%
; ii) with respect to $P$ any other MTNP $(\Identity, t)$ can take all the forms $(\Identity, u^{-1} t u)$, in particular it could be diagonal.

We remember that for any $t \in \OO{n}$ its eigenvalues are either couples of complex conjugates $e^{\pm i \theta}$ or $1$ or $-1$ and that there always exists an orthogonal matrix $E \in \OO{n}$ such that $E^T t E := \Lambda$ is quasi-diagonal, namely along the diagonal we find either $\pm 1$ or $\SO{2}$ matrices and, without loss of generality, we will assume $\Lambda$ sorted containing, from the first row, first the eigenvalues $+ 1$ and $- 1$ and then the $\SO{2}$ matrices.
\begin{MS_Proposition}
\label{condition_Witt_basis}
Given two MTNP $(\Identity, t_1)$ and $(\Identity, t_2)$ they form a Witt decomposition of $\R^{n,n}$ if and only if $\det(t_2 - t_1) \ne 0$ or, equivalently, if $t_1^T t_2$ has no eigenvalues $+1$.
\end{MS_Proposition}
\begin{proof}
The two MTNP form a Witt decomposition of $\R^{n,n}$ if and only if $\det \left(\begin{array}{c c} \Identity & \Identity \\ t_1 & t_2 \end{array}\right) \ne 0$ but by block matrices properties this determinant is $\det(t_2 - t_1)$; moreover $\det (t_2 - t_1) = \det t_1 \det (t_1^T t_2 - \Identity) \ne 0$ and since $\det t_1 = \pm 1$ we deduce that $t_1^T t_2$ has not eigenvalues $+1$.
%
%
\end{proof}

\opt{margin_notes}{\mynote{mbh.note: see log p. 724}}%
In Euclidean space $ \R^{2 n}$ given a subspace $R$ there are many linear complements but only one of them, $R_\perp$, is also an orthogonal complement such that $R \oplus R_\perp = \R^{2 n}$. In neutral space $\R^{n,n}$ given MTNP subspace $R$ again there are many linear complements but an orthogonal complement is not defined since all $R$ elements are orthogonal to themselves. The linear complement of $R$ becomes unique asking for a linear complement that is also a MTNP:
%
\begin{MS_Proposition}
\label{complementary_MTNP_unicity}
For any MTNP $R = (\Identity, t) \in \mysnqG$ there always exists one and only one complementary MTNP $R_\perp = (\Identity, -t)$ such that $R \oplus R_\perp = \R^{n,n}$.
\end{MS_Proposition}

\begin{proof}
Since $\frac{1}{2} R^T R = \Identity_n$, $R$ represents also a non degenerate subspace of dimension $n$ of Euclidean space $\R^{2 n}$. Let $(A, B)$ be the most general linear complement of $R$ in $\R^{2 n}$, then
$$
\left(\begin{array}{r r} \Identity & A \\ t & B \end{array}\right)^T \left(\begin{array}{r r} \Identity & A \\ t & B \end{array}\right) = \left(\begin{array}{c c} 2 \,\Identity & A + t^T B \\ A^T + B^T t & A^T A + B^T B \end{array}\right) \dotinformula
$$
Among the linear complements of $R$ there exists a unique orthogonal complement $R_\perp$ \cite[Proposition~9.25]{Porteous_1981} that must have $B = -t A$ and has thus the form $R_\perp := (A, -t A)$ so that in $\R^{2 n}$, with proper normalization,
\begin{equation}
\label{formula_orthogonality_R_2n}
\frac{1}{2} (R, R_\perp)^T (R, R_\perp) = \frac{1}{2} \left(\begin{array}{c c} \Identity & t^T \\ A^T & - A^T t^T \end{array}\right) \left(\begin{array}{c c} \Identity & A \\ t & -t A \end{array}\right) = \left(\begin{array}{c c} \Identity & 0 \\ 0 & A^T A \end{array}\right)
\end{equation}
that shows that $A^T A$, and thus $A$, must be full rank and so $A \in \GL{n}$. Back in $\R^{n,n}$ for the same subspaces we get
\begin{equation}
\label{formula_orthogonality_R_n_n}
\frac{1}{2} (R, R_\perp)^T \left(\begin{array}{r r} \Identity & 0 \\ 0 & - \Identity \end{array}\right) (R, R_\perp) =
\left(\begin{array}{c c} 0 & A \\ A^T & 0 \end{array}\right)
\end{equation}
and thus they form a Witt decomposition of $\R^{n,n}$ since $A$ is full rank. This establishes a bijection between the unique orthogonal complement $R_\perp$ of $R$ in $\R^{2 n}$ and Witt decomposition $R \oplus R_\perp$ of $\R^{n,n}$ that proves the unicity of MTNP $R_\perp \in \R^{n,n}$ since, should there exist another, different, MTNP $(\Identity, t')$ forming a Witt decomposition with $R$, then $(\Identity, t')$ would violate the unicity of the orthogonal complement of $R$ in $\R^{2 n}$. In both $\R^{2 n}$ and $\R^{n,n}$, $R_\perp = (A, -t A)$ with $A \in \GL{n}$ and thus by Lemma~\ref{MTNP_equivalence} $R_\perp = (\Identity, -t)$.
\opt{margin_notes}{\mynote{mbh.note: is the proof deducible from Witt cancellation theorem ?}}%
\end{proof}

\noindent In $\R^{2 n}$ the basis $\frac{1}{\sqrt{2}} (R, R_\perp)$ is orthonormal; in $\R^{n,n}$ we define a Witt basis $W$ to be \emph{orthogonal} iff
\begin{equation}
\label{formula_Witt_orthogonality_def}
W^T \left(\begin{array}{r r} \Identity & 0 \\ 0 & - \Identity \end{array}\right) W = \left(\begin{array}{r r} 0 & \Identity \\ \Identity & 0 \end{array}\right)
\end{equation}
so $\frac{1}{\sqrt{2}} (R, R_\perp)$ is orthogonal like the standard basis $\frac{1}{\sqrt{2}} \left(\begin{array}{r r} \Identity & \Identity \\ \Identity & - \Identity \end{array}\right)$, that is basis (\ref{formula_Witt_basis}) with a more traditional normalization.
\opt{margin_notes}{\mynote{mbh.note: the normalization of (\ref{formula_Witt_basis}) and (\ref{formula_P_Q_def}) is treated in logbook pp. 343 ff.}}%

We remark that the request of forming an orthogonal Witt basis in $\R^{n,n}$ is stronger than the similar request of forming an orthonormal basis in $\R^{2 n}$: by (\ref{formula_orthogonality_R_2n}) $A \in \OO{n}$ is enough in $\R^{2 n}$ but for $\R^{n,n}$ we need $A = \Identity$ (\ref{formula_orthogonality_R_n_n}).
\opt{margin_notes}{\mynote{mbh.note: rearranging the basis in either $R$ or $R_\perp$ renders the Witt basis not orthogonal}}%

\begin{MS_Corollary}
\label{coro_orthogonal_Witt_basis}
A Witt basis $W$ is orthogonal if and only if it is of the form $W = \frac{1}{\sqrt{2}} \left(\begin{array}{r r} \Identity & \Identity \\ t & -t \end{array}\right)$ for any $t \in \OO{n}$.
\end{MS_Corollary}
\begin{proof}
We already saw that $W$ is orthogonal, conversely let $\frac{1}{\sqrt{2}} \left(\begin{array}{r r} \Identity & \Identity \\ t_1 & t_2 \end{array}\right)$ be an orthogonal Witt basis then
$$
\left(\begin{array}{r r} 0 & \Identity \\ \Identity & 0 \end{array}\right) = \frac{1}{2} \left(\begin{array}{r r} \Identity & \Identity \\ t_1 & t_2 \end{array}\right)^T \left(\begin{array}{r r} \Identity & 0 \\ 0 & - \Identity \end{array}\right) \left(\begin{array}{r r} \Identity & \Identity \\ t_1 & t_2 \end{array}\right) = \frac{1}{2} \left(\begin{array}{c c} 0 & \Identity - t_1^T t_2 \\ \Identity - t_2^T t_1 & 0 \end{array}\right)
$$
and thus $\frac{1}{2} (\Identity - t_1^T t_2) = \frac{1}{2} (\Identity - t_2^T t_1) = \Identity$, namely $t_2 = - t_1$.
\end{proof}
\noindent Bijection (\ref{formula_bijection_Nn_O(n)}) thus extends also to orthogonal Witt bases of $\R^{n, n}$ by the map
\opt{margin_notes}{\mynote{mbh.note: to be compared with (\ref{formula_orthogonality_R_n_n}), see also log p. 727.2.1.2'}}%
\begin{equation}
\label{formula_Witt_bijection}
t \to \frac{1}{\sqrt{2}} \left(\begin{array}{r r} \Identity & \Identity \\ t & -t \end{array}\right) \dotinformula
\end{equation}

\section{A special property of Witt decompositions}
\label{sec_special_property}
\opt{margin_notes}{\mynote{mbh.ref: see log. pp. 718' \& 719'.}}%
Cartan first proved a crucial property of \mysnqG{} \cite[p. 96]{Cartan_1937} that only rarely surfaces in subsequent literature, see \eg \cite[Proposition~2]{BudinichP_1989}, even if, as we will see, it is tightly connected to the pivotal Cartan theorem \cite[Theorem~9.41]{Porteous_1981}. We give a constructive proof of a slightly extended result more amenable to an algorithmic exploitation needed for \SAT{} problems \cite{Budinich_2019}.

\begin{MS_Proposition}
\label{prop_2_BudinichP_1989_2}
Given any two MTNP of \mysnqG{} of $\R^{n,n}$ of forms $(\Identity, t_1)$ and $(\Identity, t_2)$ with $t_1, t_2 \in \OO{n}$ let $0 \le k \le n$ be their \emph{incidence}, namely the dimension of subspace $(\Identity, t_1) \cap (\Identity, t_2)$, then:
\opt{margin_notes}{\mynote{mbh.note: this is thus a generalization of Proposition~$13$ of paper 766 that is the particular case with $k = 0$ of this one}}%
\begin{itemize}
\item[-] it is always possible to put the two given MTNP in the respective forms $(\Identity, \Identity)$ and $(\Identity, \lambda)$ with $\lambda = \left(\begin{array}{c c} \Identity_k & 0 \\ 0 & - \Identity_{n - k} \end{array}\right) \in \O1{n}$;
\item[-] incidence $k$ is also the multiplicity of eigenvalue $+1$ in $t_1^T t_2$ and in $\lambda$.
\end{itemize}
\end{MS_Proposition}
\begin{proof}
We prove first that the incidence $k$ of $(\Identity, t_1)$ and $(\Identity, t_2)$ is also the multiplicity of eigenvalue $+ 1$ of $t_1^T t_2$. If $k = 0$ the two MTNP form a Witt decomposition of $\R^{n,n}$ and thus, by Proposition~\ref{condition_Witt_basis}, $t_1^T t_2$ has no eigenvalues $+1$. For $k > 0$ $(\Identity, t_1) \cap (\Identity, t_2) \ne \{ 0 \}$; we prove the equality showing that there is a one to one correspondence between elements of $(\Identity, t_1) \cap (\Identity, t_2)$ and the eigenvectors, of eigenvalue $+1$, of $t_1^T t_2$. Any element of $(\Identity, t_1) \cap (\Identity, t_2)$ must be of the form $(a, t_1 (a)) = (a, t_2 (a))$ that implies $t_1 (a) = t_2 (a)$ that in turn implies $t_1^T t_2 (a) = a$ and thus $a \in \R^n$ is an eigenvector of eigenvalue $+1$ of $t_1^T t_2$. Conversely each eigenvector $a$ of eigenvalue $+1$ of $t_1^T t_2$ satisfies $t_1 (a) = t_2 (a)$ and thus $(a, t_1 (a)) \in (\Identity, t_1) \cap (\Identity, t_2)$. Being $t_1^T t_2 \in \OO{n}$ a non defective matrix the equality is proved.

By Lemma~\ref{MTNP_diagonalization} $\left(\begin{array}{c c} \Identity & \Identity \\ t_1 & t_2 \end{array}\right)$ may take the form $\left(\begin{array}{c c} \Identity & \Identity \\ \Identity & t \end{array}\right)$ with $t = t_1^T t_2 \in \OO{n}$ and let $E \in \OO{n}$ be the orthogonal matrix that makes $t$ quasi-diagonal and sorted, namely $E^T t E = \Lambda$ where $\Lambda$ is a diagonal matrix containing on the diagonal first $+ 1$ and $- 1$ eigenvalues and then $\SO{2}$ elements. By Lemma~\ref{MTNP_equivalence} $(\Identity, t) \simeq (E, t E) = (E, E \Lambda)$ and given the orthogonality properties of eigenvectors in $E$ we can split $\R^n$ into two orthogonal subspaces corresponding respectively to eigenvectors of eigenvalues $+ 1$ and $\ne + 1$ of $t$. In the first subspace of dimension $k$ we get $(E', E' \Lambda') = (E', E') \simeq (\Identity_k, \Identity_k)$. For the second subspace we have $(E'', E'' \Lambda'') \simeq (\Identity_{n-k}, E'' \Lambda'' E''^T) = (\Identity_{n-k}, t'')$ with $t'' \in \OO{n-k}$ and without $+ 1$ eigenvalues and thus, by Proposition~\ref{condition_Witt_basis}, forming with $(\Identity_{n-k}, \Identity_{n-k})$ a Witt decomposition of the subspace $\R^{n - k, n - k}$ so that $(\Identity, t'') \simeq (\Identity_{n-k}, - \Identity_{n-k})$. Recomposing the two orthogonal subspaces we proved that $(\Identity, t) \simeq (\Identity, \lambda)$ with $\lambda = \left(\begin{array}{c c} \Identity_k & 0 \\ 0 & - \Identity_{n - k} \end{array}\right) \in \O1{n}$.
\end{proof}

We remark that with a different ordering of eigenvectors in $E$ then $\Lambda$ and $\lambda$ are ``unsorted'' and in particular $\lambda$ can take $\bino{n}{k}$ different forms.
We understand also the meaning of $k$: the incidence of MTNP $(\Identity, t_1)$ and $(\Identity, t_2)$ is the same in any form. This result is more exploitable in the form:
\begin{MS_Corollary}
\label{coro_t_lambda}
Given MTNP $P = (\Identity, \Identity)$ we may bring any MTNP $(\Identity, t)$ of incidence $k$ with $P$ to the form $(\Identity, \lambda_t)$ with $\lambda_t = \left(\begin{array}{c c} \Identity_k & 0 \\ 0 & - \Identity_{n - k} \end{array}\right) \in \O1{n}$.
\end{MS_Corollary}

A remarkable consequence of these results is:
\begin{MS_Corollary}
\label{coro_2^n_MTNP}
The set \mysnqG{} of neutral space $\R^{n,n}$ contains exactly $2^n$ different MTNP corresponding, via isomorphism (\ref{formula_bijection_Mn_O^n(1)}), to the elements of $\O1{n}$; all other $t \in \OO{n}$ being just another form of one of these $2^n$ MTNP.
\end{MS_Corollary}
\begin{proof}
By Proposition~\ref{isomorphism_restricted} \mysnqG{} contains at least $2^n$ different MTNP; Corollary~\ref{coro_t_lambda} shows that they can be no more than $2^n$ since given $P$ any other MTNP $(\Identity, t)$ of \mysnqG{} can take the form $(\Identity, \lambda_t)$, with $\lambda_t \in \O1{n}$.
%
%
%
\end{proof}

\noindent We remark that in any real quadratic space $V$ of dimension $2 n$ from any subspace $R$ we can build decomposition $V = R \oplus R_\perp$ and $R$ defines $R_\perp$ uniquely. But while in Euclidean space $V = \R^{2 n}$ we can choose $R$ from an infinite set, in neutral space $V = \R^{n,n}$ the situation is different: if $R$ is a MTNP its choice is limited to a discrete set of $2^n$ elements and these properties transfer, via isomorphism (\ref{formula_bijection_Nn_O(n)}), to $\OO{n}$. As a consequence the involutions of $\O1{n}$ appear as a backbone of $\OO{n}$ and this is supported by a, simple, alternative proof of the fundamental Cartan theorem \cite[Theorem~9.41]{Porteous_1981} (here restricted to the Euclidean case):
\opt{margin_notes}{\mynote{mbh.ref: see log. p. 727.2.1.2}}%
\begin{MS_theorem}
\label{Theorem_Cartan_Dieudonnè}
Any orthogonal transformation $t \in \OO{n}$ of Euclidean space $\R^n$ is expressible as the composite of at maximum $n$ hyperplane reflections.
\end{MS_theorem}
\begin{proof}
By an orthogonal base change we can put any $t \in \OO{n}$ in its quasi-diagonal form $\Lambda$. In this basis any $- 1$ eigenvalue is an hyperplane reflection and since any $\SO{2}$ element is the composite of $2$ non commuting hyperplane reflections we deduce that any $t$, with eigenvalue $+ 1$ of multiplicity $k$, is the composite of $n - k$ hyperplane reflections.
\end{proof}


\section{The different forms of MTNP $(\Identity, t)$}
\label{sec_different_forms_MTNP}
By Lemma~\ref{MTNP_diagonalization} we may freely choose any $(\Identity, t)$ to represent the reference MTNP: a very natural choice for this reference MTNP is $P = (\Identity, \Identity)$ (\ref{formula_P_Q_def}) since $t = \Identity$ is the identity of $\OO{n}$ and keeps the same form in any basis. We will adopt this choice and from now on, when we deal with a ``lonely'' MTNP $(\Identity, t)$, we will tacitly assume also ``together with $P = (\Identity, \Identity)$''. In this fashion any MTNP $(\Identity, t)$ has an implicit incidence $k$ with $P$ that, by Proposition~\ref{prop_2_BudinichP_1989_2}, is the multiplicity of the eigenvalue $+ 1$ of $t$.

Given $P$ we want to study the equivalent forms a MTNP $(\Identity, t)$ may assume and we define, for any $t \in \OO{n}$, the subset of $\OO{n}$ which elements $u$ represent the same abstract MTNP in the same basis, namely $(\Identity, u) \simeq (\Identity, t)$ a necessary condition being clearly that $u$ and $t$ have same incidence with $P$.
\opt{margin_notes}{\mynote{mbh.note: do we need to define $\OO{n} \times \OO{n} / \OO{n} \simeq \OO{n}$ ?}}%
%
We define also the Cayley chart (or transform) ${\cal C}$ \cite{Porteous_1981, Hori_Tanaka_2010}
\opt{margin_notes}{\mynote{mbh.ref: see log. p. 723'}}%
\begin{equation}
\label{formula_Cayley_chart}
{\cal C} : \End_- \R^n \to \SO{n} \quad s \to {\cal C}(s) = (\Identity - s) (\Identity + s)^{-1}
\end{equation}
where $s$ is an antisymmetric endomorphism of $\R^n$ and the map is known to be injective but not surjective since any ${\cal C}(s)$ do not have eigenvalues $- 1$ while $\SO{n}$ elements can have $- 1$ eigenvalues with even multiplicity.
\opt{margin_notes}{\mynote{mbh.note: having no eigenvalues $- 1$ is more than being in $\SO{2}$ that means having and \emph{even} number of $- 1$, see log. p. 726}}%
The Cayley chart is defined also on the larger domain of $\End \R^n$ elements that do not have eigenvalues $- 1$ and is a (nonlinear) involution since ${\cal C}^2 = \Identity$ \cite{Hori_Tanaka_2010}.

\begin{MS_Proposition}
\label{prop_Cayley_map}
Given $P = (\Identity, \Identity)$ the possible forms of MTNP $(\Identity, t)$, of incidence $k$ with $P$, and thus by Corollary~\ref{coro_t_lambda} with $\lambda = \left(\begin{array}{c c} \Identity_k & 0 \\ 0 & - \Identity_{n -k} \end{array}\right)$, are all and only those of the form $(\Identity, u)$ with $u = \lambda {\cal C}(s)$ with $s \in \End_- \R^n$ and $s = \left(\begin{array}{c c} 0_k & 0 \\ 0 & s' \end{array}\right)$ (obviously $s' \in \End_- \R^{n - k}$) and thus, indicating with $C_{\lambda} \subset \OO{n}$ the set of the equivalent forms of MTNP $(\Identity, t)$
\begin{equation}
\label{formula_C_t_general}
C_\lambda = \left\{ \lambda {\cal C}(s) : s \in \End_- \R^n, s = \left(\begin{array}{c c} 0_k & 0 \\ 0 & s' \end{array}\right) \right\} \dotinformula
\end{equation}
\end{MS_Proposition}
\begin{proof}
We prove first the two cases of incidences $k = 0$ or $n$ in a slightly more general settings, namely with respect to a generic MTNP rather than $P$. For $k = 0$ we look for those equivalent forms $u \in \OO{n}$ for $t$ such that $(\Identity, u)$ form a Witt decomposition of $\R^{n, n}$ with the unique complement $(\Identity, -t)$ and these forms are all and only those that satisfy
$$
\det \left(\begin{array}{c c} \Identity & \Identity \\ -t & u \end{array}\right) = \det (u + t) = \pm \det (t^T u + \Identity) \ne 0
$$
and thus all and only $u$ such that $t^T u$ has no eigenvalues $- 1$ and there always exists $s \in \End_- \R^n$ \cite{Porteous_1981} such that $t^T u = {\cal C}(s)$, namely $u = t {\cal C}(s)$. For incidence $k = n$ we look for those $u \in \OO{n}$ such that $(\Identity, u) \simeq (\Identity, t)$ and thus such that $u(a) = t(a)$ for any $a \in \R^n$ and the unique possibility is $u = t$.
\opt{margin_notes}{\mynote{mbh.note: also since $t^T u$ have all $+ 1$ eigenvalues namely $t^T u = \Identity$, see other proofs at log p. 727.2.1.4}}%
With respect to $P$ the possible forms of incidences $k = 0$ and $n$ are thus respectively $u = - \Identity {\cal C}(s)$ and $u = \Identity$ that proves (\ref{formula_C_t_general}) since ${\cal C}(0) = \Identity$.

For generic incidence $0 < k < n$ we can split MTNP $(\Identity, t)$ into two orthogonal subspaces of dimensions $k$ and $n-k$ that have respectively incidence $k$ and $0$ with the corresponding subspaces of $P$ and we can apply to these subspaces the previously proved cases of full and null incidence obtaining that the possible form of $u$ are $\left(\begin{array}{c c} \Identity_k & 0 \\ 0 & - {\cal C}(s') \end{array}\right)$. Form (\ref{formula_C_t_general}) comes from the observation that, by (\ref{formula_Cayley_chart}), $ {\cal C} \left(\left(\begin{array}{c c} 0_k & 0 \\ 0 & s' \end{array}\right)\right) = \left(\begin{array}{c c} \Identity_k & 0 \\ 0 & {\cal C}(s') \end{array}\right)$.
\end{proof}
We can calculate the corresponding $s \in \End_- \R^n$ for any form of the MTNP $(\Identity, t)$ exploiting Cayley chart properties: \eg let $t = \lambda {\cal C}(s)$ with $t = \left(\begin{array}{c c} \Identity_k & 0 \\ 0 & t' \end{array}\right)$ then one easily gets $- {\cal C}(s') = t'$ and thus $s' = {\cal C}(-t')$.

The Cayley chart ${\cal C}$ is injective and establishes a (nonlinear) bijection between the linear space $\End_- \R^n$ and the equivalent forms of MTNP $(\Identity, t)$ that, continuing along this road, turns out to be the tangent space to the real smooth manifold of $\OO{n}$ in $\lambda$ but we do not insist on this here. It is thus relevant to investigate the properties of the $2^n$ subsets $C_\lambda$:
\opt{margin_notes}{\mynote{mbh.ref: for a discussion of $\ne \lambda$ of same $k$ see log. p. 727}}%

\begin{MS_Proposition}
\label{prop_C_lambda_intersection}
For any $\lambda, \eta \in \O1{n}$, with $\eta \ne \lambda$, then $C_\lambda \cap C_\eta = \emptyset$.
\end{MS_Proposition}
\begin{proof}
Let us suppose that there exists $t \in C_\lambda \cap C_\eta$: if $\det \lambda = - \det \eta$ the result follows immediately from (\ref{formula_C_t_general}) since $\det {\cal C}(s) = 1$; if $\det \lambda = \det \eta$ in the $2 r > 0$ coordinates in which $\lambda$ differs from $\eta$, let $r = 1$, we would need $\Identity_2 = - {\cal C}(s')$ that is impossible since ${\cal C}(s')$ do not have eigenvalues $- 1$.
\end{proof}

\begin{MS_Corollary}
\label{prop_C_lambda_unique}
The only \O1{n}{} element contained in any $C_\lambda$ is $\lambda$
\opt{margin_notes}{\mynote{mbh.note: an independent proof is commented}}%
\end{MS_Corollary}

\begin{MS_Proposition}
\label{prop_t_in_C_lambda}
Any $t \in \OO{n}$ belongs to one of the $2^n$ subsets $C_\lambda$.
\end{MS_Proposition}
\begin{proof}
By Corollary~\ref{coro_t_lambda} any $t \in \OO{n}$ can be brought to the form $\lambda \in \O1{n}$ and by Proposition~\ref{prop_Cayley_map} there exists $s \in \End_- \R^n$ such that $t = \lambda {\cal C}(s)$.
\end{proof}

\noindent It follows that the $2^n$ subsets $C_\lambda$ of $\OO{n}$ form a partition of the group and
$$
\cup_{\lambda \in \O1{n}} \; C_\lambda = \OO{n} \dotinformula
$$
%

\section{Simple spinors and MTNP}
\label{sec_Simple_spinors_MTNP}
In section~\ref{sec_neutral_R_nn} we began with the $2^n$ simple spinors of the Fock basis of $\myClg{}{}{\R^{n,n}}$ and with them we introduced the $2^n$ MTNP of $\R^{n,n}$, now we close the circle and apply our findings on MTNP and $\OO{n}$ back to spinors exploiting the one to one correspondence between MTNP and simple spinors \cite{Cartan_1937, BudinichP_1989}; we give here an elementary proof of this bijection:
\begin{MS_Proposition}
\label{prop_simple_spinors_MTNP}
Given any element of a Fock basis $\psi \in F$ and any $v \in \R^{n,n}$
\begin{equation}
\label{formula_simple_spinor_MTNP}
v \in M(\psi) \qquad \iff \qquad v \psi = 0 \dotinformula
\end{equation}
\end{MS_Proposition}
\begin{proof}
Any $v \in M(\psi)$ (\ref{formula_MTNP_M_psi}) is the linear combination of the null vectors $x_i$ but for any of these null vectors it is simple to verify with (\ref{formula_Witt_basis_properties}) that $x_i \psi = 0$ and thus $v \psi = 0$. Conversely given any $v \in \R^{n,n}$ such that $v \psi = 0$ it follows $v^2 \psi = 0$ and since $\psi \ne 0$ and $v^2 \in \R$ necessarily $v^2 = 0$ and $v$ must be null. To prove that $v \in M(\psi)$ we observe that for any $u_1, u_2 \in M(\psi)$, $u_1$ and $u_2$ are orthogonal since from $u_1 u_2 \psi = u_2 u_1 \psi = 0$ follows $\anticomm{u_1}{u_2} = 0$. Supposing $v \notin M(\psi)$ $\my_span{v} \cup M(\psi)$ would be a totally null plane of dimension $n+1$ that is impossible and so $v \in M(\psi)$.
\end{proof}

So in \myClg{}{}{\R^{n,n}} simple spinors \emph{linearize} the quadratic relation $v^2 = 0$, characterizing MTNP $M(\psi)$, to $v \psi = 0$. This is familiar in physics where Dirac equation is the linear counterpart of quadratic Klein Gordon equation.

With Propositions~\ref{prop_2_BudinichP_1989_2} and \ref{prop_simple_spinors_MTNP} we get that for each simple spinor $\psi \in F$ then $M(\psi)$ is one of the $2^n$ MTNP of $\R^{n,n}$ uniquely identified by its $h$ signature in EFB \cite{Budinich_2016} or by an involution $\lambda \in \O1{n}$. Given the reference MTNP $P$ and its associated simple spinor $\psi_P$, such that $M(\psi_P) = P$, the vacuum spinor of physics, we can resume quite concisely the action of $\lambda \in \O1{n}$ on spinors and vectors \cite{BudinichP_1989, Budinich_2016} with
\begin{equation}
\label{formula_lambda_action}
M(\lambda(\psi_P)) := M(\psi_\lambda) = (\Identity, \lambda)
\end{equation}
and by the action of $\lambda$ we get respectively from $\psi_P$ all spinors of the Fock basis $F$ and from $P$ all MTNP of \mysnqG{}. Moreover there exists also a natural bijection between $\psi \in F$ and the subset $C_\lambda \subset \OO{n}$, the tangent space to $\OO{n}$ at $\lambda$, that, all together, form a partition of $\OO{n}$.

Spinors are elements of \myClg{}{}{\R^{n,n}}, carriers of the regular representations of the algebra, this gives them other possible geometrical interpretations.

\section{Conclusions}
\label{sec_Conclusions}
We presented several more or less known facts, from different fields, in the unifying setting of \myClg{}{}{\R^{n,n}}, hoping that they can shed some light in the intriguing relations occurring between them. We try to resume all this in a commutative diagram in which the numbers refer to formulas
$$
 \begin{tikzcd}[column sep=small]
& \psi_\lambda \in F \arrow[dl, Leftrightarrow, "(\ref{formula_simple_spinor_MTNP})" '] \arrow[dr, Leftrightarrow, "(\ref{formula_lambda_action})"] & \\
(\Identity, \lambda) \in \mysetM \arrow[Leftrightarrow, "(\ref{formula_bijection_Mn_O^n(1)})"]{rr} & & \lambda \in \O1{n}
\end{tikzcd}
$$
and moreover any $\lambda \in \O1{n}$ corresponds also to an orthogonal Witt decomposition of $\R^{n,n}$ (\ref{formula_Witt_bijection}) and to one Boolean atom, namely one of the $2^n$ assignments of $n$ Boolean variables $\rho_1, \rho_2, \ldots, \rho_n$ \cite{Budinich_2019}.

We can also resume the relations between the parent sets in a somewhat ``looser'' diagram
$$
 \begin{tikzcd}[column sep=small]
& \Psi \in S \arrow[dl, leftrightarrow] \arrow[dr, leftrightarrow] & \\
(\Identity, t) \in \mysnqG \arrow[Leftrightarrow, "(\ref{formula_bijection_Nn_O(n)})"]{rr} & & t \in \OO{n}
\end{tikzcd}
$$
where especially the relations with a generic spinor $\Psi$ deserve deeper studies.

We just remember that any $t \in \OO{n}$ is represented by either an orthogonal matrix of $\R(n)$ or by a matrix of $\R(2^n)$ corresponding respectively to the vectorial and spinorial representations of $\OO{n}$.

\newpage

\opt{x,std,AACA}{

\bibliographystyle{plain} 
\bibliography{mbh}
}
\opt{arXiv,JMP}{
%
%

%
%
}

\opt{final_notes}{
\newpage

\section{Good stuff removed from the paper}
\label{sec_Good stuff removed}
We underline that this procedure (of generating all MTNP of \mysetM{} at the beginning of Section~\ref{sec_O(n)_MTNP}) is remarkably similar to that of forming \SAT{} problems: we must choose, for each of $n$ Boolean variables, wether we take it in plain or complemented form and this similarity is no chance since there is an isomorphism between the two procedures \cite{Budinich_2019}. The other way round as any spinor can be seen as the action of a succession of creation and destruction operators applied to a vacuum state, any logical expression can be built by the action of creation and destruction operators on any initial state.
\opt{margin_notes}{\mynote{mbh.note: to be developed}}%

\myseparation

It is immediate to verify:
\begin{MS_Corollary}
\label{coro_Witt_is_on_basis}
Any orthogonal Witt basis of $\R^{n,n}$ is also an orthonormal basis of $\R^{2 n}$; conversely any orthonormal basis of $\R^{2 n}$ of the form $W = \frac{1}{\sqrt{2}} \left(\begin{array}{c c} t_1 & t_1 \\ t_2 & - t_2 \end{array}\right)$, with $t_1, t_2 \in \OO{n}$, is an orthogonal Witt basis of $\R^{n,n}$.
\end{MS_Corollary}

\myseparation

We remark also that if $a \in \R^n$ is an eigenvector of eigenvalue $+1$ of $t_1^T t_2$ then $(a, -a)$ is an element of the null space of $W$. 

\myseparation

Since ${\cal C}(s) \in \SO{n}$ all elements of $C_\lambda$ (\ref{formula_C_t_general}) have the same determinant and, given the well known structure of $\OO{n}$, it is sensible to study just its subgroup $\SO{n}$ and so we can assume $C_\lambda \subset \SO{n}$. It is interesting to study the complement of $C_\lambda$, namely the set
$$
\overline{C}_\lambda := \SO{n} - C_\lambda
$$
for which we can prove

\begin{MS_Proposition}
\label{prop_C_complement}
The elements of $\overline{C}_\lambda$ are all and only those $u \in \SO{n}$ that can be expressed as $u = \lambda x$ with $x \in \SO{n}$ that has eigenvalue $- 1$.
\end{MS_Proposition}
\begin{proof}
Any $u \in \SO{n}$ can be expressed as $u = \lambda x$ with $x = \lambda u$ and clearly $\det x = 1$. Any $x \in \SO{n}$ can have eigenvalue $- 1$ only with even multiplicity $2 r$ and if $r = 0$ then there exists $s \in \End_- \R^n$ such that $x = {\cal C}(s)$ and thus by Proposition~\ref{prop_Cayley_map} $u \in C_\lambda$. Conversely if $r > 0$ this is impossible and thus $u \in \overline{C}_\lambda$.
\end{proof}

\myseparation

Old, longer proof of Proposition~\ref{prop_2_BudinichP_1989_2} but containing Cayley map
\begin{proof}
(old) By Lemma~\ref{MTNP_diagonalization} $W = \left(\begin{array}{c c} \Identity & \Identity \\ t_1 & t_2 \end{array}\right)$ may take the form $W' = \left(\begin{array}{c c} \Identity & \Identity \\ \Identity & A \end{array}\right)$ where $A = E^T t_1^T t_2 E \in \OO{n}$ and $E \in \OO{n}$ is the orthogonal matrix that makes $A$ quasi-diagonal and sorted. Let $0 \le r \le n$ of the eigenvalues of $A$ be $\pm 1$, if $r = n$ then $A \in \O1{n}$ and the thesis is true. If $r = 0$ then all $A$ eigenvalues are different from $+ 1$ and, by Proposition~\ref{condition_Witt_basis}, $W'$ is a Witt decomposition of $\R^{n,n}$ and by Proposition~\ref{complementary_MTNP_unicity} the MTNP complement of $(\Identity, \Identity)$ is unique, it follows that necessarily $(\Identity, A) \simeq (\Identity, - \Identity)$%
\footnote{it is interesting to exhibit the actual transformation $H$ of the vectors of $A$ to $-\Identity$
$$
H = \left(\begin{array}{c c} \Identity & \Identity \\ \Identity & -\Identity \end{array}\right) W'^{-1} = \left(\begin{array}{c c} \Identity & 0 \\ - (\Identity + A) (\Identity - A)^{-1} & 2 (\Identity - A)^{-1} \end{array}\right)
$$
and $\det H = 2^n \det (\Identity - A)^{-1} \ne 0$ and thus $H \in \GL{n,n}$. The attentive reader will recognize in the term $(\Identity + A) (\Identity - A)^{-1}$ the Cayley transform of $- A$ that is properly defined since $A$ has no $\pm 1$ eigenvalues but we do not insist on this point here.}%
.
In case $0 < r < n$ then $A = \left(\begin{array}{c c} \lambda' & 0 \\ 0 & A' \end{array}\right)$ with $\lambda' \in \O1{r}$ and $A'$ is the orthogonal matrix made by $\SO{2}$ elements. Since the elements of $\lambda'$ are already $\pm 1$ to deal with $A'$ we reduce the initial problem to the case $\left(\begin{array}{c c} \Identity & \Identity \\ \Identity & A' \end{array}\right)$ in subspace $\R^{n - r, n - r}$ and we already proved that in this case $(\Identity, A') \simeq (\Identity, -\Identity)$ and, back in $\R^{n, n}$, we have thus proved that we can always bring $A = \left(\begin{array}{c c} \lambda' & 0 \\ 0 & A' \end{array}\right)$ to the form $\lambda = \left(\begin{array}{c c} \lambda' & 0 \\ 0 & - \Identity_{n - r} \end{array}\right) \in \O1{n}$. This also proves that the number $k$ of $+1$ in $\lambda$ is exactly the multiplicity of eigenvalue $+1$ of $t_1^T t_2$ that were already in $\lambda'$ and thus $\lambda = \left(\begin{array}{c c} \Identity_k & 0 \\ 0 & - \Identity_{n - k} \end{array}\right)$.

We now prove that $k$ is also the incidence of $(\Identity, t_1)$ and $(\Identity, t_2)$: if $k = 0$ the two MTNP of $W$ form a Witt decomposition of $\R^{n,n}$ and thus, by Proposition~\ref{condition_Witt_basis}, $t_1^T t_2$ has no eigenvalues $+1$. For $k > 0$ $(\Identity, t_1) \cap (\Identity, t_2) \ne \{ 0 \}$; we prove the equality showing that there is a one to one correspondence between elements of $(\Identity, t_1) \cap (\Identity, t_2)$ and the eigenvectors, of eigenvalue $+1$, of $t_1^T t_2$. Any element of $(\Identity, t_1) \cap (\Identity, t_2)$ must be of the form $(a, t_1 (a)) = (a, t_2 (a))$ that implies $t_1 (a) = t_2 (a)$ that in turn implies $t_1^T t_2 (a) = a$ and thus $a \in \R^n$ is an eigenvector of eigenvalue $+1$ of $t_1^T t_2$. Conversely each eigenvector $a$ of eigenvalue $+1$ of $t_1^T t_2$ satisfies $t_1 (a) = t_2 (a)$ and thus $(a, t_1 (a)) \in (\Identity, t_1) \cap (\Identity, t_2)$. Being $t_1^T t_2 \in \OO{n}$ a non defective matrix the equality is proved.
\end{proof}
%

The proof was in two steps: in the first we performed an orthogonal basis change that brought MTNP $(\Identity, t_1^T t_2)$ to its quasi-diagonal, sorted form, in this step the eigenvalues of $t_1^T t_2$ did not change, as in any change of basis. In the second step to get the desired form $(\Identity, \lambda)$ we applied a non orthogonal rotation to the vectors spanning the MTNP, in this step the eigenvalues of $t_1^T t_2$ did change: all the complex conjugate eigenvalues went to $-1$.


} 

\end{document}